\documentclass[preprint,12pt]{elsarticle}

\usepackage{amssymb}
\usepackage{amsmath}
\journal{Journal of Subatomic Particles and Cosmology}

\begin{document}

\newcommand{\ttbar}{\mathrm{t}\overline{\mathrm{t}}}
\newcommand{\ttg}{\ttbar\gamma}
\newcommand{\ttW}{\ttbar\mathrm{W}}
\newcommand{\ttj}{\ttbar j}

\begin{frontmatter}



\title{Measurements of top quark asymmetries}


\author[l1]{Nils Faltermann} %% Author name
\author{on behalf of the ATLAS and CMS Collaborations\fnref{copyright}}
\fntext[copyright]{Copyright 2026 CERN for the benefit of the ATLAS and CMS Collaborations. Reproduction of this article or parts of it is allowed as specified in the CC-BY-4.0 license}

\affiliation[l1]{organization={Karlsruhe Institute of Technology},%Department and Organization
            %% addressline={Kaiserstraße 12},
            city={Karlsruhe},
            %% postcode={76131},
            %% state={},
            country={Germany}}

%% %% Abstract
\begin{abstract}
  The study of top quark asymmetries at the LHC provides an excellent opportunity to probe subtle differences in the production of top quarks and antiquarks made by the standard model of particle physics. In this contribution, the latest experimental results on this topic by the ATLAS and CMS Collaborations are summarized.

\end{abstract}

%% Keywords
%% \begin{keyword}
%% %% keywords here, in the form: keyword \sep keyword

%% %% PACS codes here, in the form: \PACS code \sep code

%% %% MSC codes here, in the form: \MSC code \sep code
%% %% or \MSC[2008] code \sep code (2000 is the default)

%% \end{keyword}

\end{frontmatter}


\section{Introduction}\label{sec:into}
With several hundred million top quarks produced up to this date, it's no wonder the LHC is called a top quark factory. However, precision of results in the top quark sector become more and more limited by systematic uncertainties, both on the experimental and theoretical side. The measurement of asymmetries may represent a solution to this problem, where systematic uncertainties cancel out as they affect both top quarks and antiquarks in a similar way. Not only does this allow to probe standard model (SM) predictions made by higher-order quantum chromodynamics (QCD), but also to test for potential beyond the SM (BSM) contributions in the context of effective field theory (EFT) approaches. In the following, the latest results from the ATLAS~\cite{ATLAS:2008xda} and CMS~\cite{CMS:2008xjf,CMS:2023gfb} experiments at the LHC, with proton-proton at a center-of-mass energy of $\sqrt{s} = 13\,\mathrm{TeV}$, are presented.

\section{Asymmetry in top quark pair production}\label{sec:ttbar}
The Feynman diagrams at leading order in QCD contributing to the production of top quark pairs ($\ttbar$) do not introduce an asymmetry between the top quark and antiquark at the LHC. Only at higher orders, e.g. when considering quark-gluon initial states or quark-quark box diagrams, a small difference in the rapidity distributions between top quarks and antiquarks arises. Contrary to the Tevatron, where the asymmetry resulted in a relative shift of the rapidity distribution because of proton-antiproton collisions, the asymmetry at the LHC manifests itself in a broader rapidity distributions for top quarks compared to antiquarks. The so-called \textit{charge asymmetry} is defined as
\begin{align*}
  A_{\mathrm{C}}^{\ttbar} = \frac{N(\Delta|y_{\ttbar}| > 0) - N(\Delta|y_{\ttbar}| < 0)}{N(\Delta|y_{\ttbar}| > 0) + N(\Delta|y_{\ttbar}| < 0)},
\end{align*}
with the number of events $N$ in a given data sample and $\Delta|y_{\ttbar}| = |y_{\mathrm{t}}| - |y_{\overline{\mathrm{t}}}|$ as the rapidity difference between the top quark ($\mathrm{t}$) and the antiquark ($\overline{\mathrm{t}}$).

\begin{figure}[h]
\centering
\includegraphics[width=.4\linewidth]{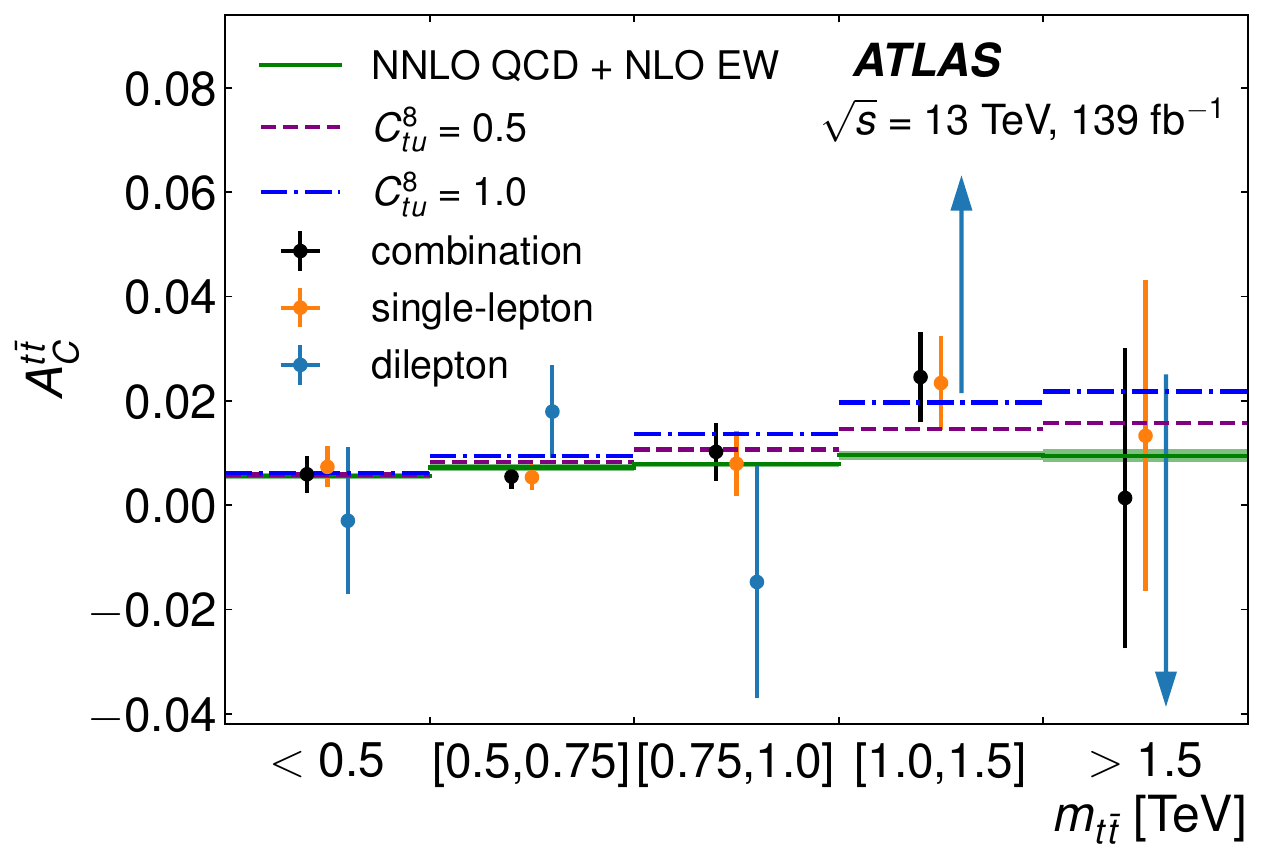}
\includegraphics[width=.4\linewidth]{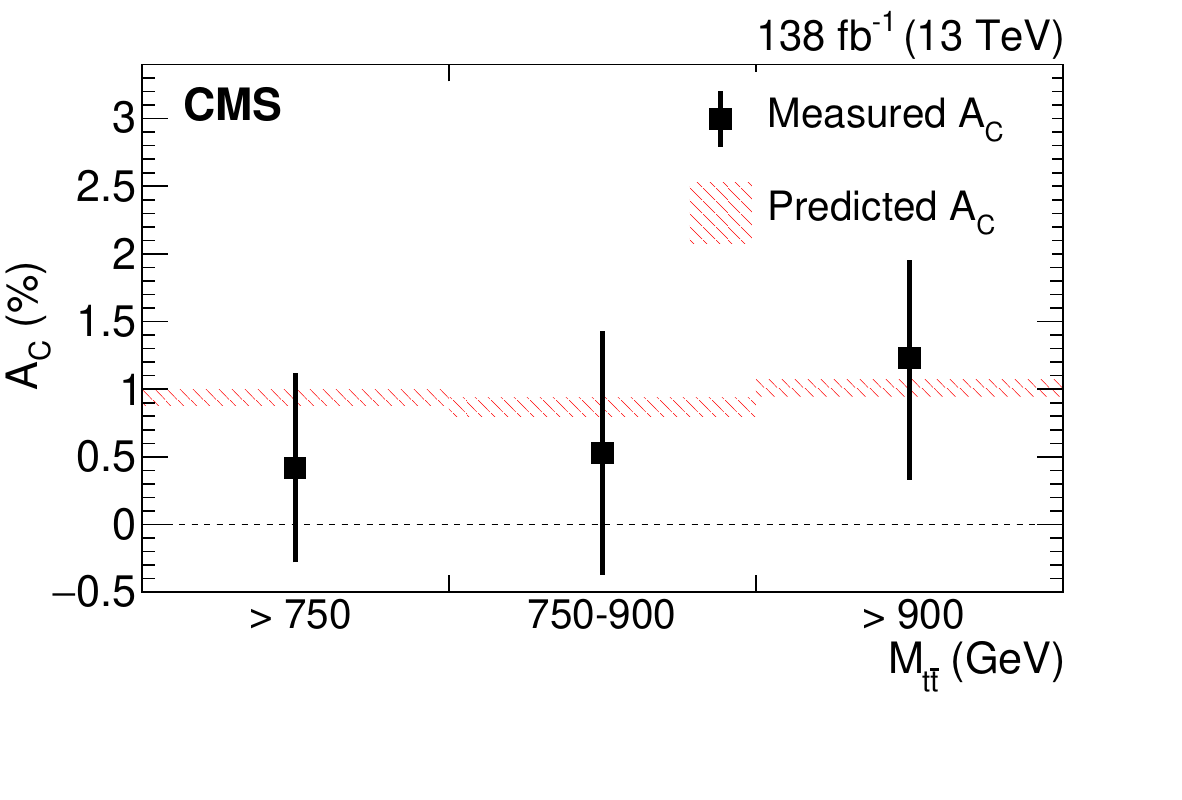}
\caption{Results of the differential measurement of the top quark charge asymmetry from the ATLAS (left) and CMS (right) Collaborations~\cite{ATLAS:2022waa,CMS:2022ged}.}\label{fig:ttbar}
\end{figure}

The ATLAS Collaboration has measured the charge asymmetry individually in the lepton+jets and dilepton final states of the $\ttbar$ system, as well as combined in both channels~\cite{ATLAS:2022waa}. With a Fully Bayesian Unfolding technique an inclusive value for the charge asymmetry at parton level of
\begin{align*}
  A_{\mathrm{C}}^{\ttbar}(\mathrm{exp.}) = 0.0068 \pm 0.0015
\end{align*}
was obtained, which is compatible with the SM prediction of $A_{\mathrm{C}}^{\ttbar}(\mathrm{theo.}) = 0.0064 ^{+0.0005} _{-0.0006}$ and results in a significance of $4.7\sigma$ against the no-asymmetry scenario. This is the first evidence for the top quark charge asymmetry at the LHC. In addition, the charge asymmetry has also been measured differentially and compared both to the SM prediction and BSM scenarios with different EFT operators, as depicted exemplary for the invariant mass of the $\ttbar$ system in Fig.~\ref{fig:ttbar} (left).
No significant deviations from the SM prediction was observed.

The CMS Collaboration has measured the charge asymmetry in $\ttbar$ production as well, but focuses only on boosted topologies in the lepton+jets channel for an enhanced BSM sensitivity~\cite{CMS:2022ged}. No significant deviations with respect to the SM prediction and the no-asymmetry scenario was observed. The differential result with respect to the invariant mass of the $\ttbar$ system is shown in Fig.~\ref{fig:ttbar} (right).

\section{Top quark pair production in association with vector bosons}\label{sec:ttgttW}
The abundance of gluon-gluon initial states for $\ttbar$ production is limiting the magnitude of the top quark charge asymmetry at the LHC. A way to enhance the asymmetry is to require the presence of an isolated photon in the final state, which could arise from initial-state radiation of quarks in the initial state. While increasing the charge asymmetry in this associated production of $\ttbar$ with a photon ($\ttg$), the overall production cross section becomes much lower and the photon introduces ambiguity as it may as well be radiated off from any charged decay product of the top quark.

\begin{figure}[h]
\centering
\includegraphics[width=.3\linewidth]{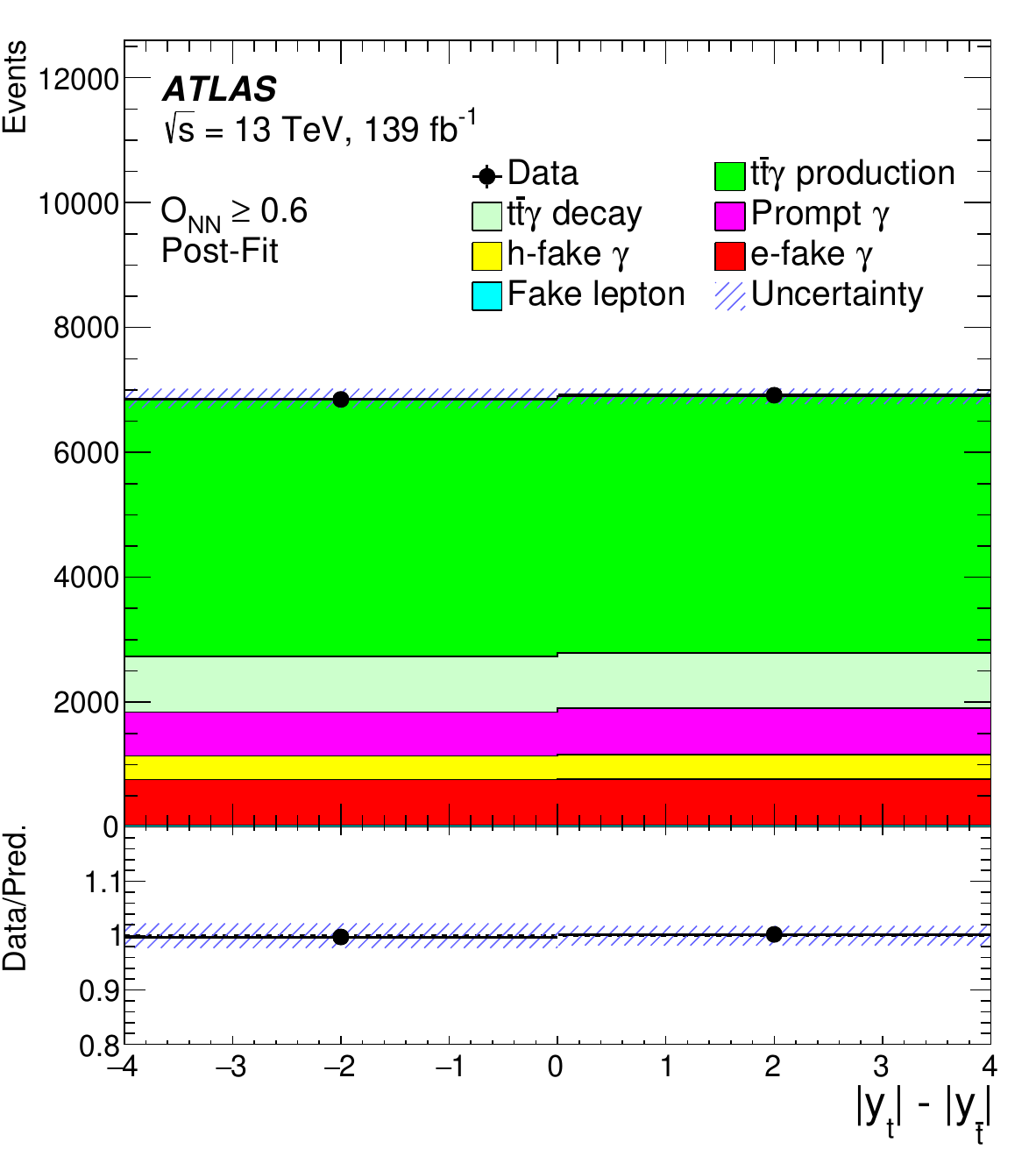}
\includegraphics[width=.35\linewidth]{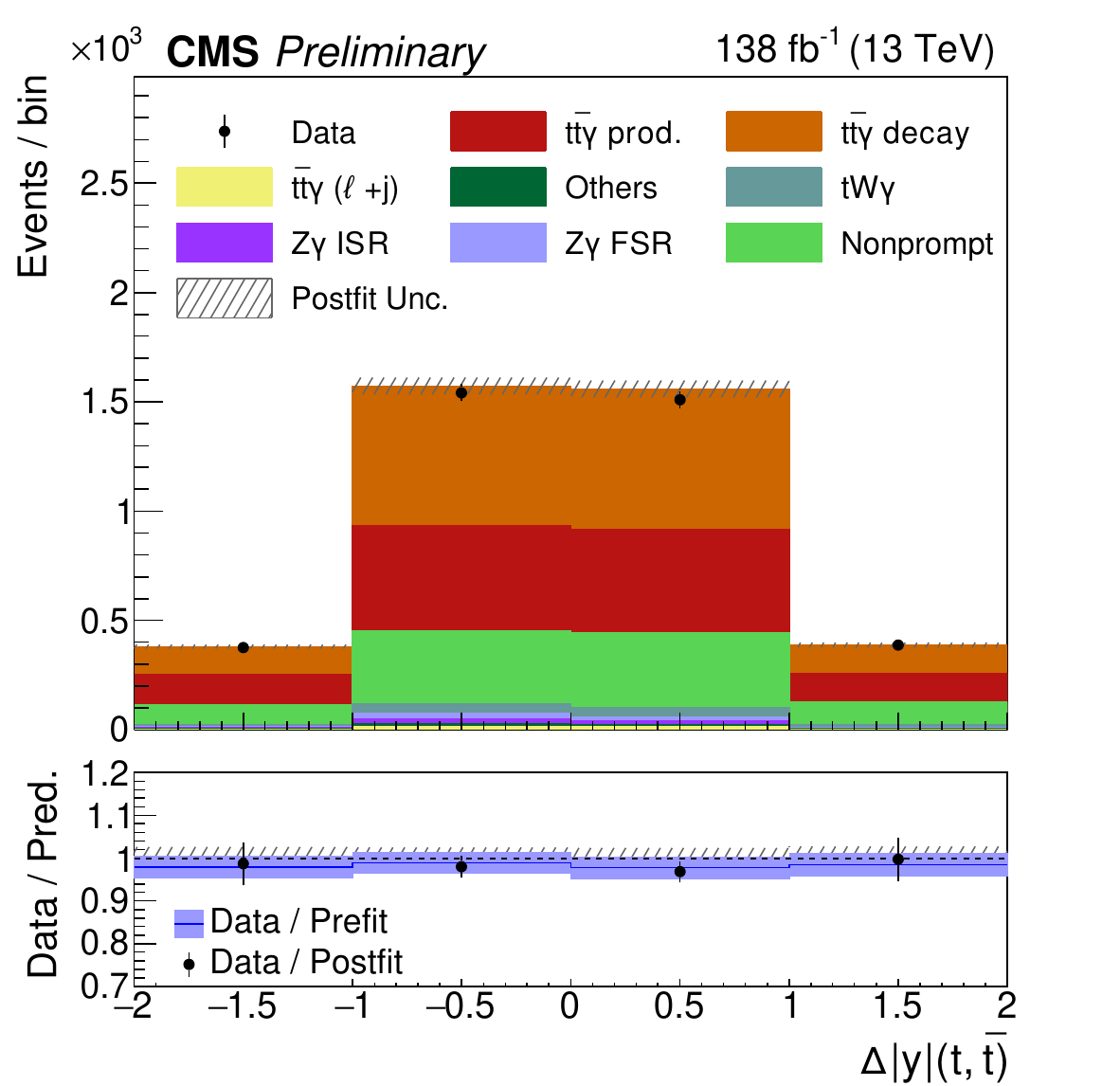}
\caption{Results for the top quark charge asymmetry in $\ttg$ production by the ATLAS (left) and CMS (right) experiments~\cite{ATLAS:2022wec,CMS:2025oki}.}\label{fig:ttg}
\end{figure}

Both the ATLAS and CMS Collaborations have studied this production mode and measured the charge asymmetry, with the former focusing on the lepton+jets final state~\cite{ATLAS:2022wec} and the latter on the dilepton~\cite{CMS:2025oki} final state of the $\ttbar$ system. The results for both measurements are shown in Fig.~\ref{fig:ttg}.
The measured values of $A_C^{\mathrm{ATLAS}} = -0.003 \pm 0.029$ and $A_C^{\mathrm{CMS}} = (-1.2 \pm 4.2)\%$ are both compatible with no asymmetry and the theoretical predictions in their respective phase-space regions.

A similar approach can be made if the photon is replaced by a W boson, resulting in a charge asymmetry in associated $\ttW$ production. With the additional W boson, which is required to decay leptonically, the assignment of leptons in the final state to the two W bosons from the top quark decays, which are also required to decay leptonically, and the third W boson becomes nontrivial. For this reason, the asymmetry in this channel is not defined by the rapidity difference between the top quarks, but instead by the rapidity difference of the leptons from the subsequent decay of the top quarks.

\begin{figure}[h]
\centering
\includegraphics[width=.525\linewidth]{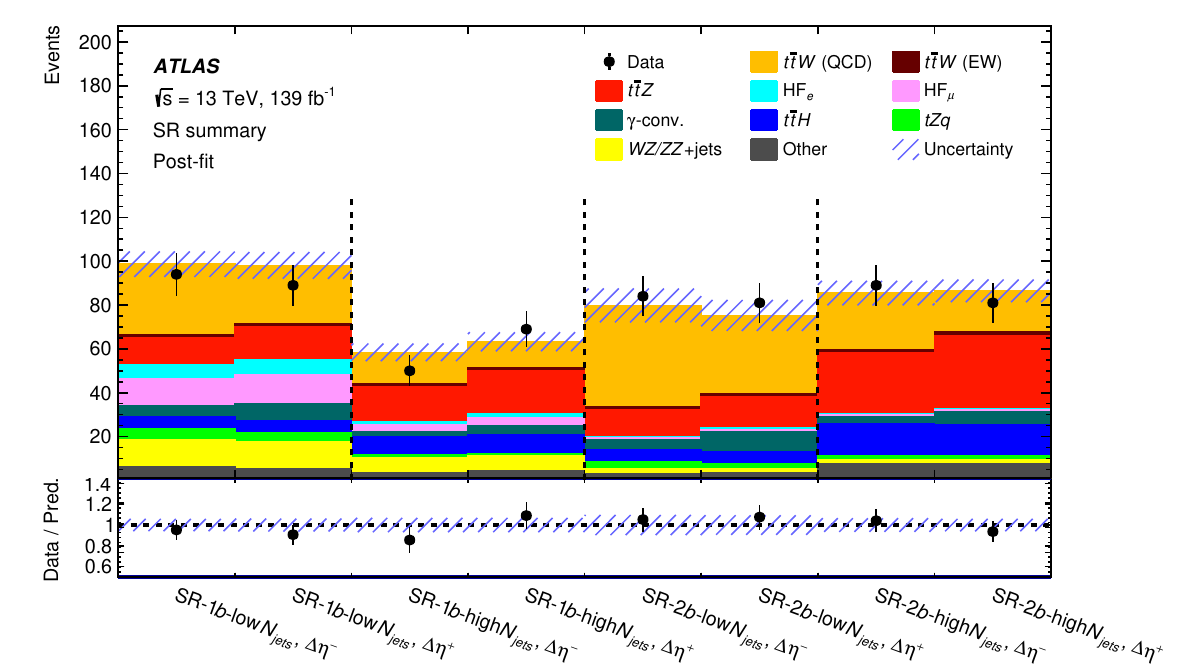}
\includegraphics[width=.315\linewidth]{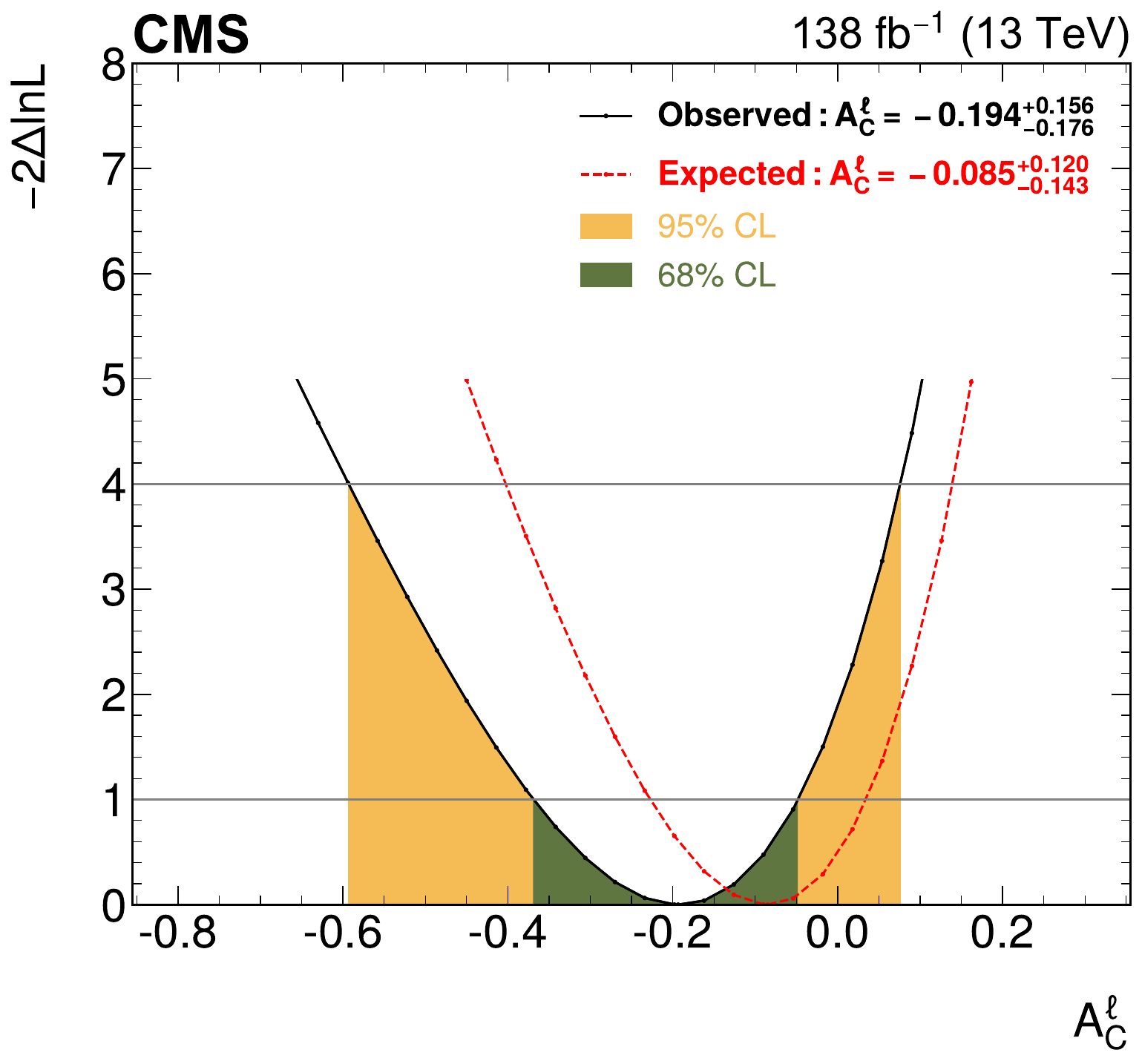}
\caption{Lepton charge asymmetry results obtained by the ATLAS~\cite{ATLAS:2023xay} (left) and CMS~\cite{CMS:2025iwa} (right) Collaborations in associated $\ttW$ production.}\label{fig:ttW}
\end{figure}

Results are available both by the ATLAS~\cite{ATLAS:2023xay} and CMS~\cite{CMS:2025iwa} experiments in the three lepton final state as shown in Fig.~\ref{fig:ttW}.
While the results by the ATLAS Collaboration at reconstruction and particle level are compatible with no asymmetry and the theoretical predictions, the result by the CMS Collaboration of $A_C^{l} = -0.19 ^{+0.16} _{-0.18}$ shows a deviation from zero by a little more than 1$\sigma$.

\section{Energy and incline asymmetry in top quark pair plus jet production}\label{sec:ttj}
In addition to the top quark charge asymmetry, one can also define the so-called \textit{energy asymmetry} of top quarks in associated $\ttbar$ plus jet production ($\ttj$):
\begin{align*}
  A_E (\theta_j) = \frac{\sigma^{\mathrm{opt}}(\theta_j | \Delta E > 0) - \sigma^{\mathrm{opt}}(\theta_j | \Delta E < 0)}{\sigma^{\mathrm{opt}}(\theta_j | \Delta E > 0) + \sigma^{\mathrm{opt}}(\theta_j | \Delta E < 0)},
\end{align*}
with $\sigma^{\mathrm{opt}} = \sigma(\theta_j | y_{\ttj} > 0) + \sigma(\pi - \theta_j | y_{\ttj} < 0)$, where $\theta_j$ is the scattering angle between the initial partons and the additional jet in the final state. A related quantity is the \textit{incline asymmetry}, defined as:
\begin{align*}
A_I = \frac{\sigma_A^{\varphi}(y_{\ttj} > 0) - \sigma_A^{\varphi}(y_{\ttj} < 0)}{\sigma_A^{\varphi}(y_{\ttj} > 0) + \sigma_A^{\varphi}(y_{\ttj} < 0)},
\end{align*}
with $\sigma_A^{\varphi} = \sigma(\cos \varphi > 0) - \sigma(\cos \varphi < 0)$, where $\varphi$ is the angle between the planes defined by the initial- and final-state momenta of the $\ttj$ system.

\begin{figure}[h]
\centering
\includegraphics[width=.38\linewidth]{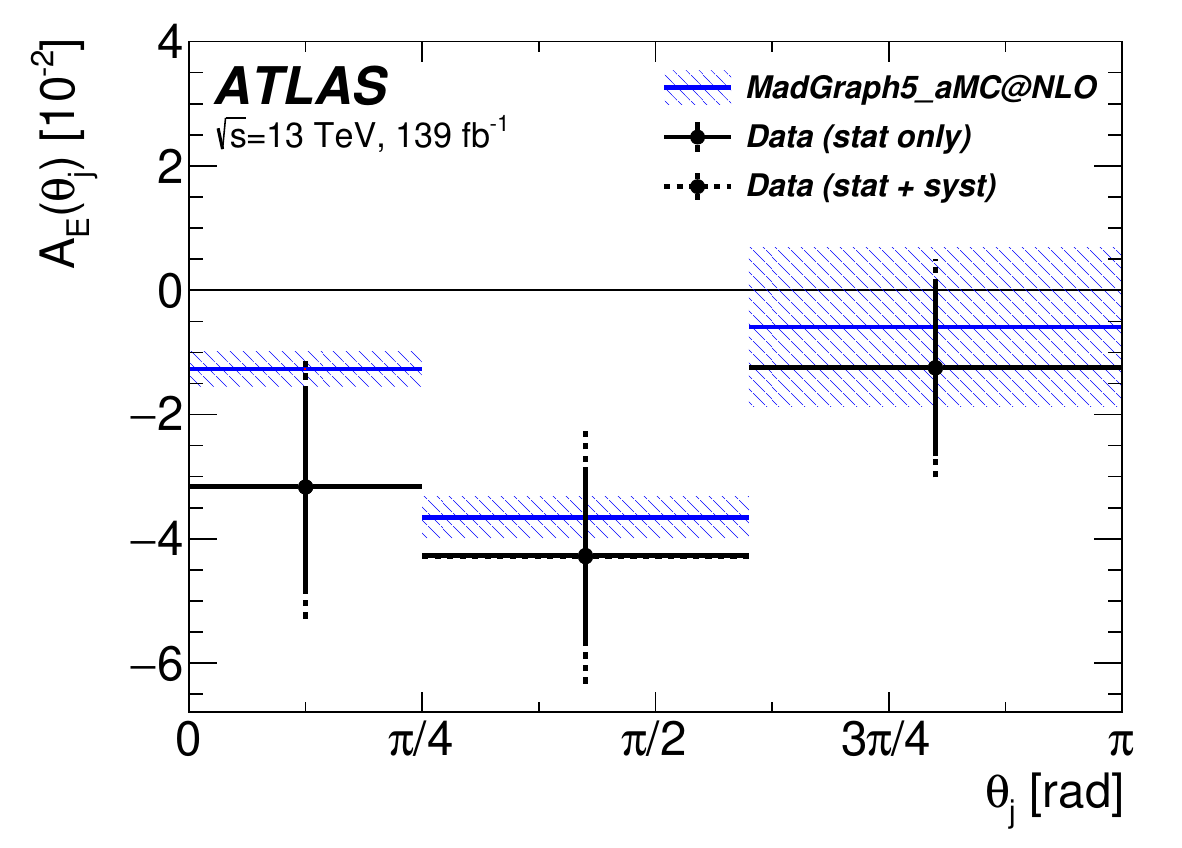}
\includegraphics[width=.475\linewidth]{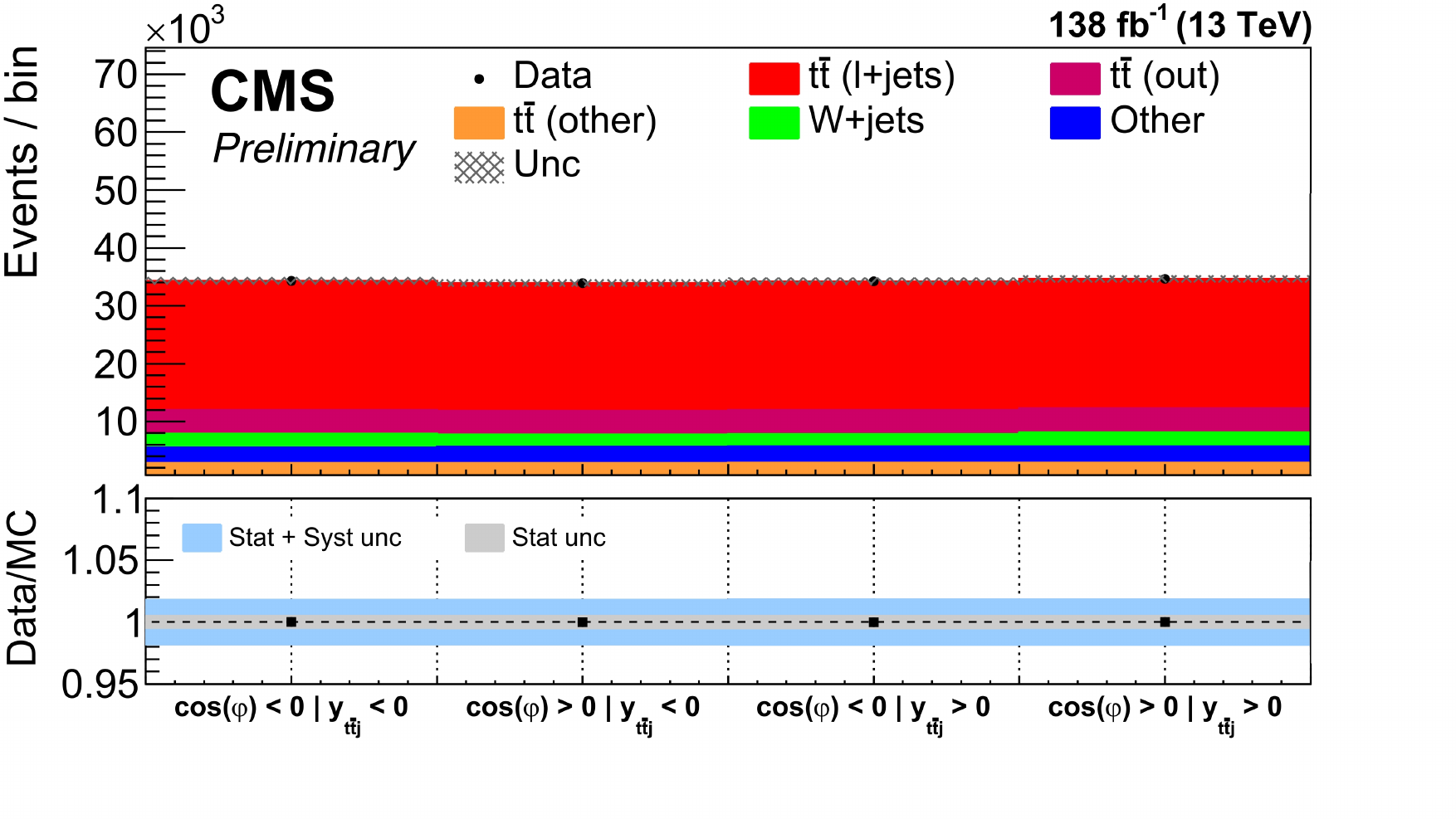}
\caption{Exemplary results for the energy asymmetry in $\ttj$ production by the ATLAS Collaboration~\cite{ATLAS:2021dqb} (left) and for the incline asymmetry by the CMS Collaboration~\cite{CMS:2025qmd} (right).}\label{fig:ttj}
\end{figure}

Measurements of the energy asymmetry have been carried out by both the ATLAS~\cite{ATLAS:2021dqb} and CMS~\cite{CMS:2025qmd} Collaborations, while the latter also includes a result for the incline asymmetry as well. The results are exemplary depicted in Fig.~\ref{fig:ttj}.
For the energy asymmetry result of the ATLAS experiment no significant deviations from the theoretical predictions are observed, with a difference from zero of $2.1\sigma$ in the most significant bin. Similarly, the CMS experiment finds a significance of $2.7\sigma$ against no asymmetry in the most significant bin ($A_E = (-6.3 \pm 2.3)\%$), but also a deviation of $2\sigma$ with respect to the theoretical predictions~\cite{CMS:2025qmd}. For the incline asymmetry a value of $A_I = (2.5 \pm 2.3)\%$ is measured, which although less significant, still deviates from zero and the slightly negative theoretical predictions~\cite{CMS:2025qmd}.

\section{Summary}
Top quark asymmetries provide an excellent way to test precision predictions in the top quark sector at the LHC. While results for inclusive top quark pair production become limited by systematic effects, the sensitivity for results in association with vector bosons or jets is still limited by the amount of data recorded and will be improved in the future.




\end{document}